\documentclass[twocolumn,prl,aps,floatfix,epsf,psfig,groupaddress,amssymb]{revtex4}
\usepackage{graphicx,subfigure}
\usepackage{color,soul}
\usepackage{epsfig}
\usepackage{mathtools}
\DeclarePairedDelimiter{\evdel}{\langle}{\rangle}
\newcommand{\ev}{\operatorname{}\evdel}
\usepackage{amsmath}
\pdfoutput=1
\usepackage{float}
\usepackage{hyperref}

\begin{document}

\title{Effects of random atomic disorder on the magnetic
  stability of graphene nanoribbons with zigzag edges}

\author{K. E. \c{C}akmak}
\author{A. Alt{\i}nta\c{s}}
\author{A. D. G\"u\c{c}l\"u}

\affiliation{Department of Physics, Izmir Institute of Technology,
  IZTECH, TR35430, Izmir, Turkey}

\date{\today}

\begin{abstract}
We investigate the effects of randomly distributed atomic defects on
the magnetic properties of graphene nanoribbons with zigzag edges
using an extended mean-field Hubbard model. For a balanced defect
distribution among the sublattices of the honeycomb lattice in the
bulk region of the ribbon, the ground state antiferromagnetism of the
edge states remains unaffected. By analyzing the excitation spectrum,
we show that while the antiferromagnetic ground state is susceptible
to single spin flip excitations from edge states to magnetic defect
states at low defect concentrations, it's overall stability is
enhanced with respect to the ferromagnetic phase.
\end{abstract}
\maketitle

%\section{Introduction}

The possibility to induce magnetism in graphene through sublattice
engineering of the honeycomb lattice can potentially lead to a new
class of spintronic and magnetic nanodevices\cite{Awschalom+13,
  Wolf+01,Chappert+07,Son+06,Wang+09,Fernández+07,Yazyev+08,
  Wimmer+08,Trauzettel+07,GUCLU+15,GUCLU+09}. Indeed, although pure
graphene is not expected to be magnetic, Lieb's bipartite lattice
theorem for Hubbard model\cite{Liebs-11} predicts a finite total spin
related to breaking of the sublattice symmetry. This broken symmetry
can happen, for instance, at the zigzag edges of a graphene
nanostructure\cite{fujita+96,HosikLee+05,YWSon+06,Li+08,Cervantes+08,
Tobias+08,GUCLU+13,Ulas+16,Sahin+10,Yazyev+10,Jung+09,
Potasz+10,HFeldner+10,Magda+14,Modarresi+17} or around an atomic defect
\cite{Yazyev+07,Palacios+08,Jaskolski+15,Gonzalez+16,Gargiulo+14,
  Esquinazi+03,Soriano+10,Singh+09,GucluBulut2015,Zhang+16,Yazyev2+08,Pereira+06}, 
  resulting in magnetized localized states.

In zigzag graphene nanoribbons (ZGNR), as the opposite edge atoms
belong to opposite sublattices, one expects antiferromagnetically
coupled zigzag localized edge states with zero total spin.  The
induced magnetic behavior is predicted by several theoretical models,
including density functional theory (DFT) \cite{HosikLee+05,YWSon+06,GUCLU+09,
Soriano+10,Singh+09}, the mean-field approximation of Hubbard model
\cite{fujita+96,Palacios+08,Yazyev2+08,Ulas+16,Jung+09,Yazyev+10},
exact diagonalization\cite{GUCLU+09,Modarresi+17} and quantum Monte Carlo 
simulation\cite{HFeldner+10}. However, on the experimental side\cite{Lepold+17,
Giang+17,Chong+18,Dimas+16,Hayashi+17,Robert+17,Jordan+17,Ayrat+18,Nataliya+17,
Chen+17,Jacobse+17}, the direct observation of magnetism in graphene 
nanoribbons is still lacking, most likely due to limited control over edge 
structure. Recently, a semiconductor-to-metal transition as a function of 
ribbon width was observed in nanotailored graphene ribbons with zigzag edges 
\cite{Magda+14}, attributed to a magnetic phase transition from the 
antiferromagnetic (AFM) configuration to the ferromagnetic (FM) configuration, 
raising hopes for the fabrication of graphene-based spintronic and magnetic 
storage devices. Possible theoretical explanations for the observed AFM to 
FM transition in ZGNR include doping\cite{Schubert+12,Topsakal+08,Dai+13} 
and formation of electron-hole puddles through long range Coulomb 
impurities\cite{Ulas+16}.

Atomic defects have also a significant influence on the magnetic
properties of graphene, as was shown before in several theoretical
work\cite{Yazyev+07,Palacios+08,Jaskolski+15,Gonzalez+16,Gargiulo+14,
  Esquinazi+03,Soriano+10,Singh+09,GucluBulut2015,Zhang+16}.
Recently, the existence of magnetism in graphene by using hydrogen
atoms was observed \cite{Gonzalez+16} and another direct experimental
evidence of the magnetism in graphene due to single atomic vacancy in
graphene was detected by using scanning tunnelling
microscope\cite{Zhang+16}. An open question is the effect of the
induced magnetic moment by a random distribution of atomic defects on
the stability of the antiferromagnetic phase of the ZGNR.

In this work, we investigate the magnetic phases of ZGNRs containing
10010 atoms with randomly distributed atomic defects using mean-field
Hubbard calculations. We show that the atomic defects stabilizes the
antiferromagnetic phase of the ZGNR.  Our finding suggests that it
should be easier to directly observe and control magnetism in ZGNRs
through a generation of randomly distributed atomic defects (vacancies
or adatoms) on the bulk region of the ribbon.

Our starting point is the tight-binding model for $p_z$ orbitals,
where $s$, $p_x$, and $p_y$ orbitals are disregarded as they mainly
contribute to the mechanical stability of graphene. Atomic defects are
modeled as randomly distributed vacancies, where the $p_z$ orbitals
are simply removed from the honeycomb lattice. The vacancy mimics the
hybridization of the corresponding $p_z$ orbital with a hydrogen
adatom. Lattice distortion effects due to hydrogenation are neglected
and zigzag edge atoms are taken to be free of defects, assuming
a controlled hydrogenation of nanoribbon's bulk region only. We
consider defect concentrations of 1-5\% of the total number of
atoms. Within the extended Hubbard model, the Hamiltonian is given by

\begin{align}
H_{MFH} =& \sum_{ij\sigma} ( t_{ij} c^{\dagger}_{i\sigma} c_{j\sigma} + h.c) 
\nonumber 
\\
&+ U\sum_{i} \left(\ev{n_{i\uparrow}} - \frac{1}{2}\right)n_{i\downarrow} + 
\left(\ev{n_{i
\downarrow}} - \frac{1}{2}\right)n_{i\uparrow} \nonumber \\
&+ \sum_{ij} V_{ij} \left(\ev{n_{j}-1}n_{i\downarrow} + \ev{n_{j}-1}n_{i
\downarrow}
\right)
\end{align}

%Figure1
\begin{figure*}
\includegraphics[scale=0.19]{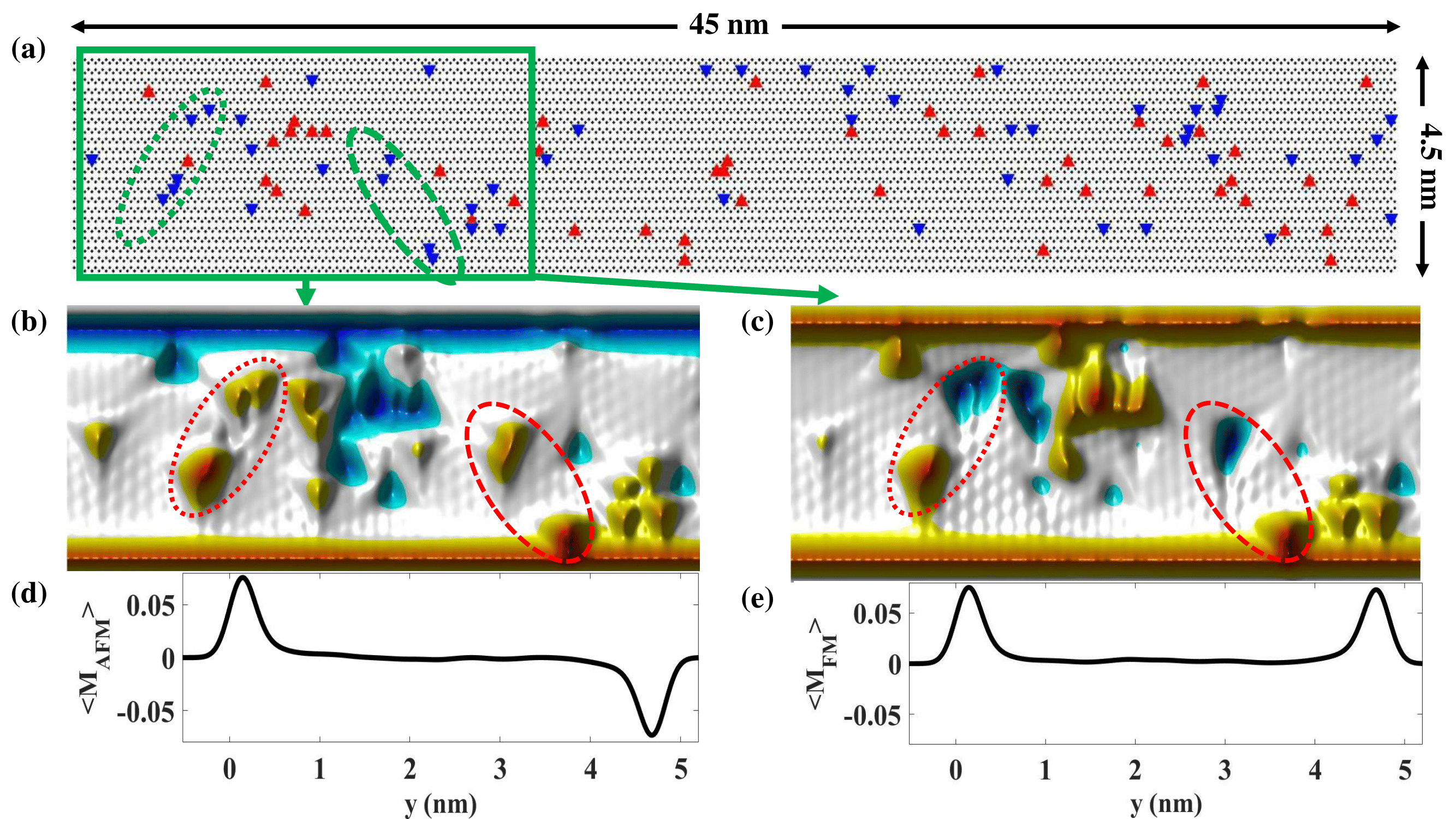}
\caption{(Color online) (a) Graphene nanoribbon lattice structure with
  10010 atoms and randomly generated 1\% defects concentration equally
  distributed within A-sublattice (downward pointed triangles) and
  B-sublattice (upward pointing triangles). (b) AFM (ground state)
  and (c) FM spin denstiy profile of a portion of the ribbon.
  (d) and (e) Average magnetization along the ribbon length for the AFM
  and FM phases.}
\end{figure*}

The first term is for the tight-binding (TB) approximation where the
$t_{ij}$ are the hopping parameters are taken to be $t_{nn}$= -2.8 eV
for the nearest neighbours and $t_{nnn}$= -0.1 eV for the second
nearest neighbors\cite{Neto+09,Reich+02}. The creation
$c^{\dagger}_{i\sigma}$ and $c_{i\sigma}$ annihilation operators
create and annihilate an electron at the {\it i}th orbital with spin
$\sigma$, respectively. The terms $\ev{n_{i \sigma}}$ correspond to
the expectation value of electron densities.  The second and third
terms are on-site and long-range Coulomb interaction
terms, respectively. The on-site Coulomb potential $U$ is taken to be
$16.522/ \kappa$ eV, where $\kappa=6$ is an effective dielectric
constant. The long-range interaction parameters $V_{ij}$ are taken to
be $8.64/ \kappa$ eV and $5.33/ \kappa$ eV for the first two
neighbours, and $1/d_{ij} \kappa$ for distant
neighbors\cite{Slatermatrix}. However, unlike for ZGNRs in the
presence of long range disorder\cite{Ulas+16}, the effect of
long-range Coulomb interactions is found to be negligible in the
presence of atomic defects considered in current work.

We consider 45 nm long and 4.5 nm wide ZGNRs consisting of 10010 atoms
with various defect configurations. Figure 1a shows a ZGNR
configuration with 1\% of defects that are randomly and equally
distributed among the two sublattices of the honeycomb lattice.  The
downward pointing (blue color online) and upward pointing (red color 
online) triangles correspond to sublattice A-site and B-site vacancies, 
respectively. The self-consistent Hubbard calculations were performed 
within different $S_z$=($n_{\uparrow}$ - $n_{\downarrow}$)/2 subspaces 
to find the overall ground state. As one may suspect a competition 
between the AFM and FM states\cite{Ulas+16}, we have scanned the 
$0$ $\leq$ $S_z$ $\leq$ $130$ values, with a focus around AFM state 
$S_z$=0 and FM state $S_z=N_{edge}/2$ where the number of edge states 
is given by $N_{edge}=138$ for the clean structure. For each value 
$S_z$ the self consistent calculations were repeated with different 
initial density matrices to ensure the convergence to the global energy 
minimum.

Figures 1b and 1c shows the spin densities of a portion of the ribbon,
for the lowest energy AFM and FM states, respectively. Despite the
inclusion of long range electron interactions and second nearest
neighbour hoppings, the mean-field solution to the Hubbard model leads
to $S_z=0$ ground state in all our calculations with equally
distributed defects among the two sublattices, in agreement with
Lieb's theorem. Indeed, in Fig.1b, the A-site and B-site defects lead
to spin-up (red color online) and spin-down (blue color online)
magnetic moments, respectively, as expected. On the other hand, the
spin density distribution for the lowest FM state is harder to predict
since it is not a ground state consistent with Lieb's theorem.
Interestingly, the edge ferromagnetism of the $S_z=73$ state remains
robust (see Fig.1c) and the bulk atoms have a zero average
magnetization as shown in Fig.1d-e. This simple observation has an 
important consequence on the stability of the AFM phase with respect 
to the FM phase: For the FM phase, the magnetization of the defects 
nearby edge atoms is strictly dictated by the strong magnetization 
of the edges, locally obeying Lieb's theorem. Hence, far from the edges,
one must encounter sublattice spin frustrations where Lieb's theorem
cannot be locally satisfied, costing energy. For instance, for the AFM
state where the Lieb's theorem is globally satisfied, the A-site
defects in the encircled areas in Figs.1b are ferromagnetically
coupled to each other, whereas their coupling is antiferromagnetic in
Fig.1c. Our calculations show that such local violation of Lieb's
theorem only occurs among defect sites and never between an edge and a
defect site.

%Figure2
\begin{figure}
\includegraphics[scale=0.195]{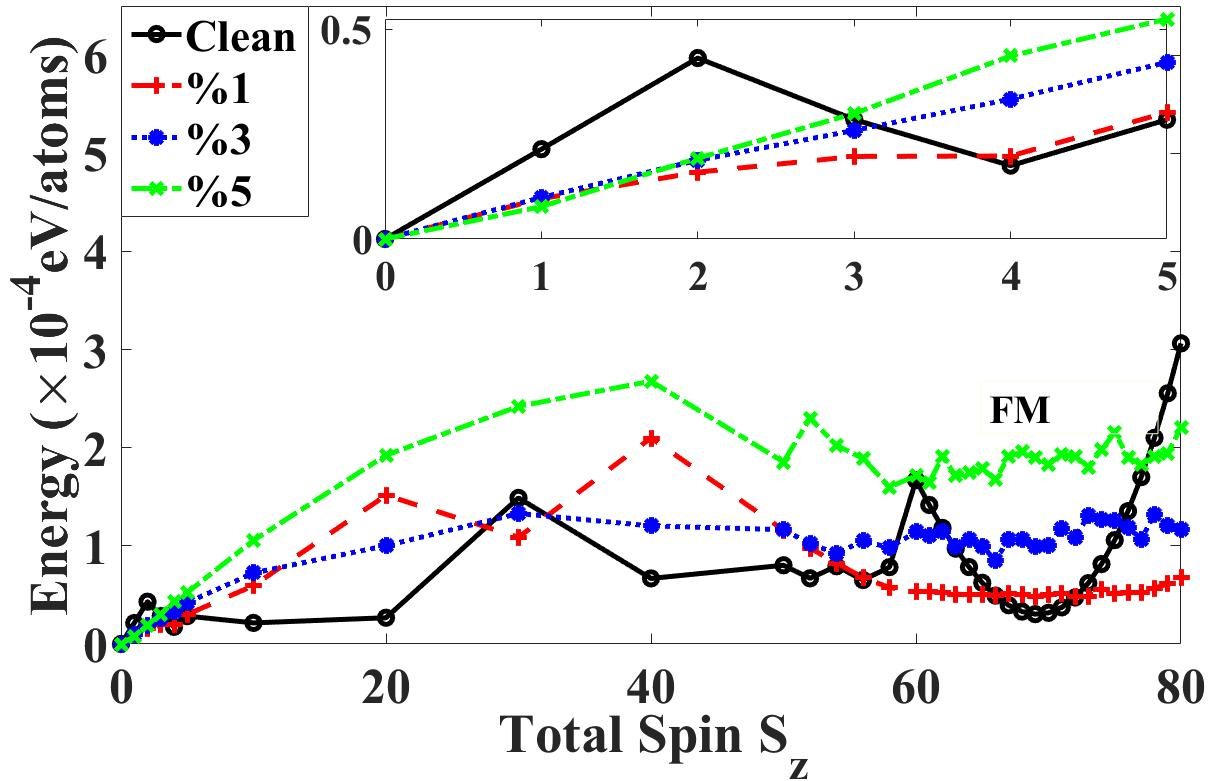}
\caption{(Color online) Mean-field energy per atom as a function of
  total spin $S_{z}$ for clean, 1\%, 3\%, and 5\% of defect
  concentrations. For the clean case, the ground state is AFM edges
  with $S_{z}$=0, and the FM phase occurs at $S_{z}$=69. The FM-AFM
  gap increases with increasing defects concentration.}
\end{figure}

As discussed above, local violation of Lieb's theorem in the bulk
region of the FM phase costs energy. A striking consequence of the
energy cost is an increased stability of the AFM phase with respect to
the FM phase. Figure 2 shows the energy per atom of different magnetic
states $S_z$ with respect to the AFM ground state, for various defect
concentrations up to $5\%$. For clean structure, the FM phase is at
$S_{z}=N_{edge}/2=69$ and the FM-AFM gap is $3.041\times10^{-5}$ eV/atom. 
As the defect concentration is increased, the FM-AFM gap increases, 
reaching $1.6\times10^{-4}$ eV for the 5\% of defects. Strikingly, the 
gap increase with respect to the AFM phase occurs not only for FM phase
but most other $S_z$ states. However, in the vicinity of $S_z=0$ (see
the inset), i.e. for single/few spin flips, energy cost is decreased
at $1\%$ of defect concentrations, but then increases slightly with
increasing number of defects. This reflects the fact that for low
defect concentrations it is easier to flip an edge spin by moving it
into a defect state than into the opposite edge. We note that similar
behaviors were observed for other randomly generated defect
concentrations and a statistical analysis will be presented below in
Fig.4.

%%Figure3
\begin{figure}
\includegraphics[scale=0.5]{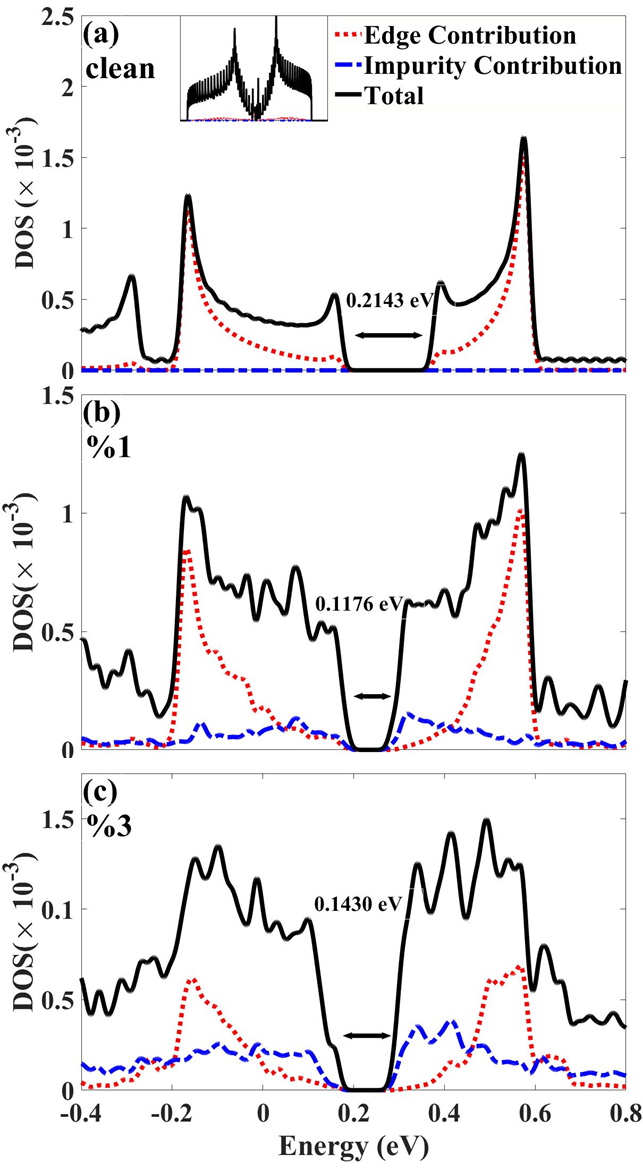}
\caption{(Color online) Mean-field DOS for the AFM phase for the (a)
  clean, (b) 1\% concentration, and (c) 3\% concentration cases. Edge
  and defect states contributions are plotted with dotted and
  dot-dashed lines. Energy gap values of the total DOS are given for
  each case.}
\end{figure}

Figure 3 shows the mean-field density of states (DOS) for AFM ground
state, for different concentrations considered in Fig.2.  The solid
lines represent the total DOS, whereas the dotted and dashed lines
represent the contribution from edge and defect atoms (more precisely,
atoms neighbouring the defects/vacancies) to the DOS, respectively. For
the clean nanoribbon, the AFM gap is 0.2143 eV, which roughly
corresponds to the energy required to flip a single spin. As the
defect concentration is increased to 1\%, there is an increase of
midgap state density and the AFM gap is decreased to 0.1176 eV. This
is consistent with the single spin flips in the vicinity of $S_z=0$
discussed in Fig.2. When the concentration of defects is increased to
3\%, the AFM gap now increases slightly. This change of behavior
reflects the fact that for higher number of defects the magnetic
coupling between the defects is enhanced in average, stabilizing the
magnetic configuration and making the spin flips harder. However, we
note that, the AFM-FM gap monotonically increases with increasing
midgap states due to the local violation of Lieb's theorem, as
discussed earlier.

%Figure4
\begin{figure}
\includegraphics[scale=0.44]{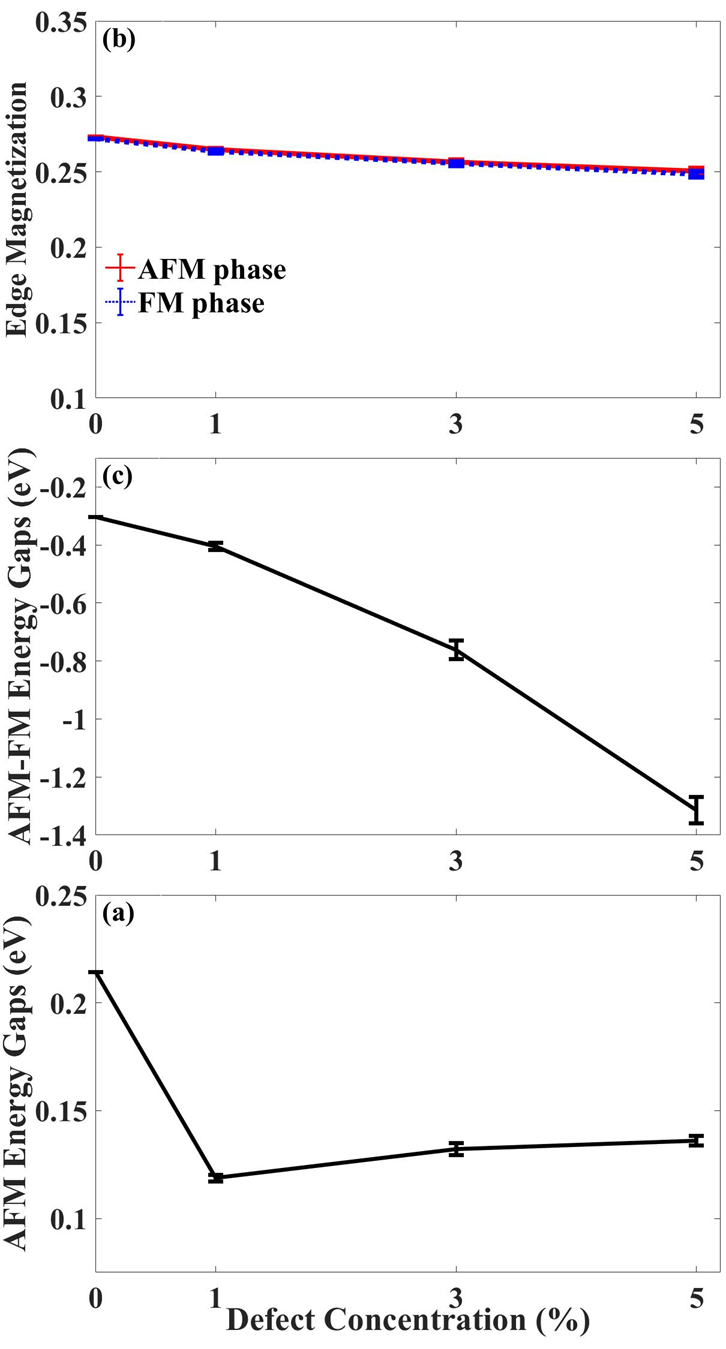}
\caption{(Color online) Average (a) single edge 
  magnetization,(b) AFM-FM energy gap, and (c) AFM energy gap, 
  over 10 randomly generated disorder configurations, as a function 
  of defect concentration.}
\end{figure}

Up to this point, the results presented were obtained for particular
randomly generated defect configurations. For a statistical analysis
of our results, we have repeated our calculations for 10 randomly
generated configuration at 1, 3, and 5\% defect concentrations. We
have observed similar behavior in all the disorder configurations and
the results are presented in Fig.4 as a function of defect
concentration.  The average magnetization of edge atoms for the AFM
and FM phases, shown in Fig.4a, decreases slightly with increasing
defect concentration.  The difference between the AFM and FM edge
magnetization remains negligible (within the error bars), consistent
with Figs.1d-e. On the other hand, Fig. 4b shows that 
the AFM-FM gap rapidly decreases in average with a small error bar, clearly
demonstrating an increased stability of the AFM phase with respect to
the FM phase. Finally, the average AFM gap shown in Fig.4c, indicating
the energy cost for a single spin flip, systematically undergoes a
decrease at lower concentrations, then keeps slowly increasing at
concentrations higher than 1\% due to a more stable magnetic lattice
formed by defects.

In summary, effects of randomly distributed atomic defects on the
stability of magnetic phases of a zigzag edged graphene nanoribbon
were investigated using a mean-field Hubbard approach. For an equal
distribution of atomic defects among the two sublattices of the
honeycomb lattice, the ground state remains antiferromagnetic with
$S_z=0$. At lower defect concentrations ( $\leq 1\%$ ), the energy of
single spin flips from the antiferromagnetic ground state is decreased
due to possible electron transfer from edges to defect
states. However, we show that the AFM-FM energy gap remains well
protected and is enhanced as a function defect concentration.  The
increased stability of the AFM-FM gap by controlling defect
concentrations opens up new possibilities for spintronic and magnetic
nanodevice applications.

{\it Acknowledgment}. This work was supported by The Scientific and
Technological Research Council of Turkey (TUBITAK) under the 1001 Grant
Project No. 114F331. The numerical calculations reported in
this work were partially performed at TUBITAK ULAKBIM, High
Performance and Grid Computing Center (TRUBA resources).

%==========================================================================
%                      REFERENCES
%==========================================================================

\end{document}